\documentclass[preprint]{aastex}

\usepackage{graphicx}
\usepackage{array}
\usepackage{natbib}
\usepackage{amssymb}
\usepackage{lscape}

\begin{document}

\title{Discovery of Five Candidate Analogs for $\eta$\,Carinae in Nearby Galaxies}

\author{Rubab~Khan\altaffilmark{1,2,5},
Scott~M.~Adams\altaffilmark{3},
K.~Z.~Stanek\altaffilmark{3,4},
C.~S.~Kochanek\altaffilmark{3,4},
G.~Sonneborn\altaffilmark{2}
}

\altaffiltext{1}{NASA Goddard Space Flight Center, MC 665,
8800 Greenbelt Road, Greenbelt, MD 20771; rubab.m.khan, george.sonneborn-1@nasa.gov}

\altaffiltext{2}{NASA Postdoctoral Program, ORAU,
P.O. Box 117, MS 36, Oak Ridge, TN 37831}

\altaffiltext{3}{Dept. of Astronomy, The Ohio State University, 140
W. 18th Ave., Columbus, OH 43210; sadams, kstanek, ckochanek@astronomy.ohio-state.edu}

\altaffiltext{4}{Center for Cosmology and AstroParticle Physics, 
The Ohio State University, 191 W.\ Woodruff Ave., Columbus, OH 43210}

\altaffiltext{5}{JWST Fellow}

\shorttitle{Discovery of $\eta$\,Carinae Analogs}

\shortauthors{Khan et al. 2015(c)}

\begin{abstract}
\label{sec:abstract}
The late-stage evolution of very massive stars such as $\eta$\,Carinae may be 
dominated by episodic mass ejections which may later lead to Type\,II 
superluminous supernova (SLSN-II; e.g., SN\,2006gy). However, as long as 
$\eta$\,Car is one of a kind, it is nearly impossible to quantitatively 
evaluate these possibilities. Here we announce the discovery of five objects 
in the nearby ($\sim4-8$\,Mpc) massive star-forming galaxies M\,51, M\,83, 
M\,101 and NGC\,6946 that have optical through mid-IR photometric properties 
consistent with the hitherto unique $\eta$\,Car. The \textit{Spitzer} mid-IR 
spectral energy distributions of these $L_{bol}\simeq3-8\times10^{6}\,L_\odot$ 
objects rise steeply in the $3.6-8\,\micron$ bands, then turn over between $8$ 
and $24\,\micron$, indicating the presence of warm ($\sim400-600$\,K) 
circumstellar dust. Their optical counterparts in HST images are 
$\sim1.5-2$\,dex fainter than their mid-IR peaks and require the presence of 
$\sim5-10\,M_\odot$ of obscuring material. Our finding implies that the rate of 
$\eta$\,Car-like events is a fraction $f=0.094$ ($0.040 < f < 0.21$ at 90\% 
confidence) of the core-collapse supernova (ccSN) rate. If there is only one 
eruption mechanism and SLSN-II are due to ccSN occurring inside these 
dense shells, then the ejection mechanism is likely associated with the onset of carbon 
burning ($\sim 10^3 - 10^4$~years) which is also consistent with the apparent 
ages of massive Galactic shells. 
\end{abstract} 
\keywords{stars: evolution, massive, mass-loss
--- stars: individual ($\eta$ Carinae)}
\maketitle

\section{Introduction}
\label{sec:introduction}

The last stages of the evolution of the most massive (M$\gtrsim30 $\,M$_\odot$) stars may be dominated by episodic 
large mass-ejections~\citep[e.g.,][]{ref:Humphreys_1984,ref:Smith_2014}. This leads to dust condensing out of the ejecta, 
obscuring the star in the optical but revealing it in the mid-infrared (mid-IR) as the absorbed UV and optical photons 
are re-emitted at longer wavelengths~\citep[e.g.,][]{ref:Kochanek_2012a}. 
The best known example is $\eta$\,Carinae ($\eta$\,Car) which contains one of 
the most massive (100-150\,$M_\odot$) and most luminous ($\sim5\times10^{6}\,L_\odot$) 
stars in our Galaxy~\citep[e.g.,][]{ref:Robinson_1973}. Its Great Eruption in the mid-1800s 
led to the ejection of $\sim10 M_{\odot}$ of material~\citep{ref:Smith_2003} now seen as a dusty nebula around 
the star. While ongoing studies are helping us further analyze the Great Eruption
\citep[see, e.g.,][]{ref:Rest_2012,ref:Prieto_2014},
deciphering the rate of such events and their consequences is challenging because 
no analog of this extraordinary laboratory for stellar astrophysics 
(in terms of stellar mass, luminosity, ejecta mass, time since mass ejection etc.) has previously been found. 

A related puzzle is the existence of the Type\,II superluminous supernovae (SLSN-II) that are plausibly 
explained by the SN ejecta colliding with a massive shell of previously ejected material~\citep[e.g., SN\,2006gy;][]{ref:Smith_2007b}.
A number of SNe, such as the Type\,Ib SN\,2006jc \citep{ref:Pastorello_2007} and
the Type\,IIn SN\,2009ip \citep[e.g.,][]{ref:Mauerhan_2012,ref:Prieto_2012,ref:Pastorello_2013}, 
have also shown transients that could be associated with mass ejections shortly prior to 
the final explosion. But the
relationship between these transients and $\eta$\,Car or other LBVs surrounded by still older,
massive dusty shells~\citep[e.g.,][]{ref:Smith_2006} is unclear. 

There are presently no clear prescriptions for how to include events like the Great Eruption into theoretical models.
Even basic assumptions --- such as whether the mass loss is triggered by the final
post-carbon ignition phase as suggested statistically by~\citet{ref:Kochanek_2012a} or by an opacity phase-transition in the 
photosphere~\citep[e.g.,][]{ref:Vink_1999}
or by interactions with a binary companion~\citep[e.g.,][]{ref:Soker_2005} 
--- are uncertain. Studies of possible mass-loss mechanisms~\citep[e.g.,][]{ref:Shiode_2014}
are unfortunately non-prescriptive on either rate or outcome.
Observationally, we are limited by 
the small numbers of high mass stars 
in this short evolutionary phase and 
searching for them in the Galaxy is complicated by having to look
through the crowded, dusty disk and distance uncertainties. Obtaining a 
better understanding of this phase 
of evolution requires exploring other galaxies.

We demonstrated in~\citet{ref:Khan_2010,ref:Khan_2011,ref:Khan_2013} that 
searching for extragalactic self-obscured stars utilizing \textit{Spitzer} images 
is feasible, and in \citet{ref:Khan_2015} we
isolated an emerging class of 18 candidate self-obscured stars with
$L_{bol}\sim10^{5.5-6.0}L_\odot$ ($M_{ZAMS}\simeq25$-$60 M_\odot$) in 
galaxies at $\sim1-4$\,Mpc. We have now expanded 
our search to the large star-forming galaxies M\,51, M\,83, M\,101 and NGC\,6946 
(distance\,$\simeq4-8$\,Mpc). We picked these galaxies 
because they have high star formation rates (total $SFR_{H\alpha}\simeq6.9 M_\odot / $yr,
mainly based on \citealp{ref:Kennicutt_2008}) and hosted 
significant numbers of core-collapse supernovae (ccSNe) over the past century~\citep[total 20, e.g.,][]{ref:Botticella_2012}, 
indicating that they are likely to host a significant number of evolved high mass stars. 

In this letter, we announce the discovery of five objects in these galaxies that have optical through 
mid-IR photometric properties consistent with the hitherto unique 
$\eta$\,Car as it is presently observed. In what follows, we  
describe our search method (Section\,\ref{sec:data}), 
analyze the physical properties of the five potential $\eta$\,Car analogs (Section\,\ref{sec:analysis})
and consider the implications of our findings (Section\,\ref{sec:discussion}). 

\section{The $\eta$\,Car Analog Candidates}
\label{sec:data}

At extragalactic distances, an $\eta$\,Car analog would appear 
as a bright, red point-source in \textit{Spitzer} IRAC~\citep{ref:Fazio_2004} images, with a 
fainter optical counterpart due to self-obscuration. 
Given enough absorption, the optical counterpart could be undetectable.
Building on our previous 
work~\citep{ref:Khan_2011,ref:Khan_2013,ref:Khan_2015,ref:Khan_2015b},
we relied on these properties to 
identify the $\eta$\,Car analog candidates.
For M\,51~\citep[D\,$\simeq8$\,Mpc,][]{ref:Ferrarese_2000}, 
M\,83~\citep[D\,$\simeq4.61$\,Mpc,][]{ref:Saha_2006} and 
M\,101~\citep[D\,$\simeq6.43$\,Mpc,][]{ref:Shappee_2011}
we used the full \textit{Spitzer} mosaics available from 
the Local Volume Legacy Survey~\citep[LVL,][]{ref:Dale_2009},
and for NGC\,6946~\citep[D\,$\simeq5.7$\,Mpc,][]{ref:Sahu_2006}
we used those from the \textit{Spitzer} Infrared Nearby Galaxies Survey 
\citep[SINGS,][]{ref:Kennicutt_2003}.
 
We built Vega-calibrated IRAC\,$3.6-8\,\micron$ and MIPS~\citep{ref:Rieke_2004} 
$24\,\micron$ point-source catalogs for each galaxy
following the procedures described in~\citet{ref:Khan_2015b}.
We use PSF photometry at $3.6$ and $4.5\,\micron$, 
a combination of PSF and aperture photometry (preferring PSF) at $5.8\,\micron$, 
and only aperture photometry 
at $8.0$ and $24\,\micron$ as the PSF size and PAH emission both increase toward 
longer wavelengths. For all sources, we 
determine the spectral energy distribution (SED) slope 
$a$~($\lambda L_\lambda \propto \lambda^a$),
the total IRAC luminosity ($L_{mIR}$) and 
the fraction $f$ of $L_{mIR}$ that is emitted in the first three IRAC bands. 
Following the selection criteria established in~\citet{ref:Khan_2013} ---
$L_{mIR}>10^{5}\,$\,L$_\odot$, $a>0$ and $f>0.3$ --- we initially selected $\sim700$ sources 
from our mid-IR point-source catalogs.

We examined the IRAC images to exclude the sources
associated with saturated, resolved or foreground objects, 
and utilized the VizieR\footnote{http://vizier.u-strasbg.fr/} web-service to 
rule out spectroscopically confirmed non-stellar sources and those with
high proper motions.
We inspected the $3.6-24\,\micron$ SEDs of the remaining sources to identify the ones that 
most closely resemble the SED of $\eta$\,Car and then queried the 
Hubble Source Catalog (HSC\footnote{https://archive.stsci.edu/hst/hsc/search.php}, Version\,1)
to exclude those with bright optical counterparts 
(m$\lesssim20$\,mag, implying $L_{opt}\gtrsim1.5-6\times10^5\,L_\odot$). These steps produced a short-list of 
$\sim20$ sources for which we retrieved archival HST images 
and the associated photometry from the Hubble Legacy 
Archive (HLA\footnote{http://hla.stsci.edu/}). Since the HST and \textit{Spitzer}
images sometimes have significant ($\sim1\farcs0$) astrometric mismatches, we utilized the
IRAF GEOMAP and GEOXYTRAN tasks to locally align the HST and \textit{Spitzer}
images with uncertainties $\lesssim0\farcs1$. We then searched for the closest 
optical counterpart within a matching radius of $0\farcs3$.

\begin{figure}[p]
\begin{center}
\begin{tabular}{c}
\includegraphics[width=80mm]{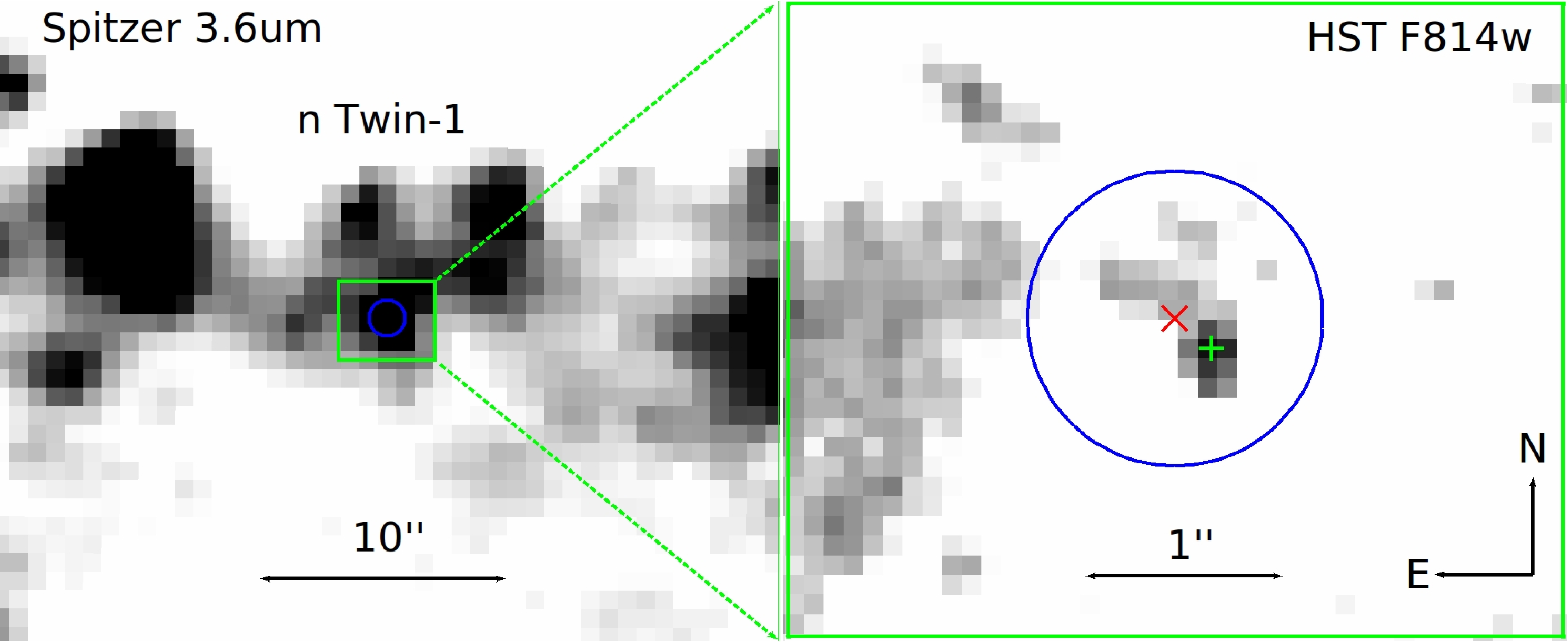} \\
\includegraphics[width=80mm]{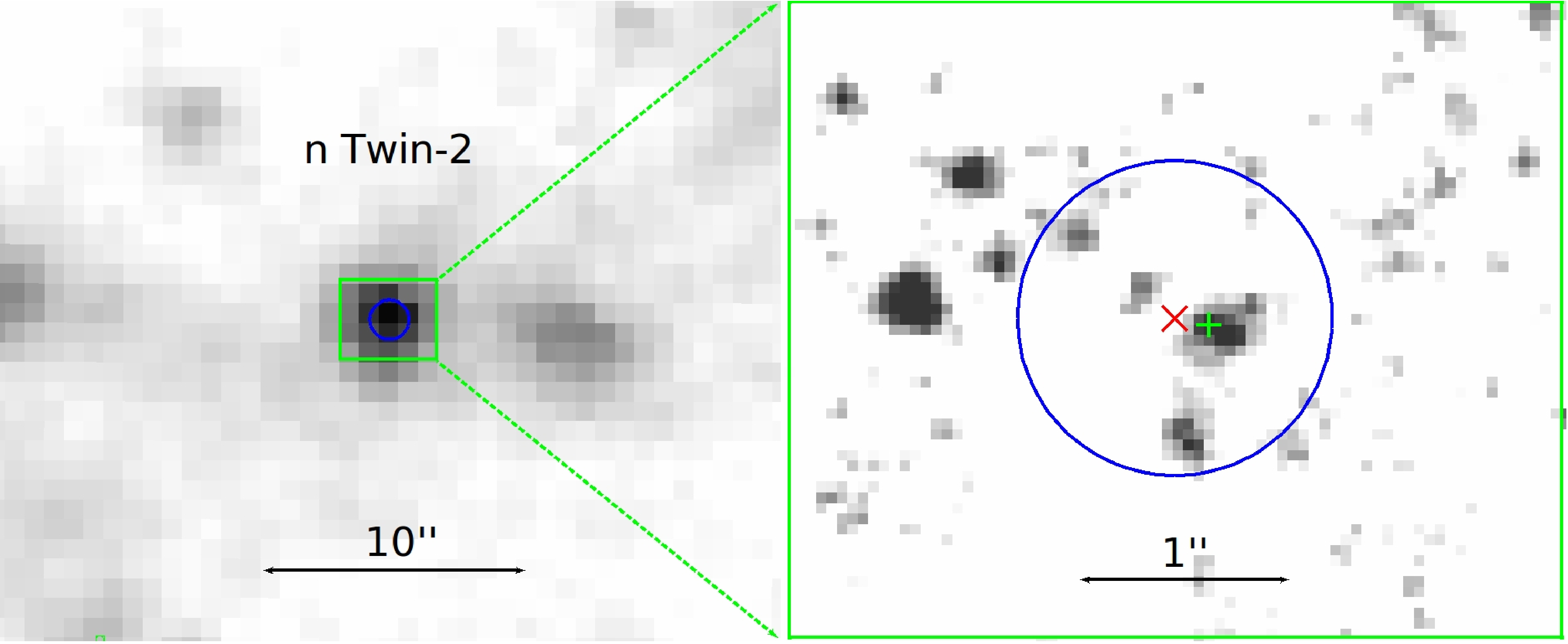} \\
\includegraphics[width=80mm]{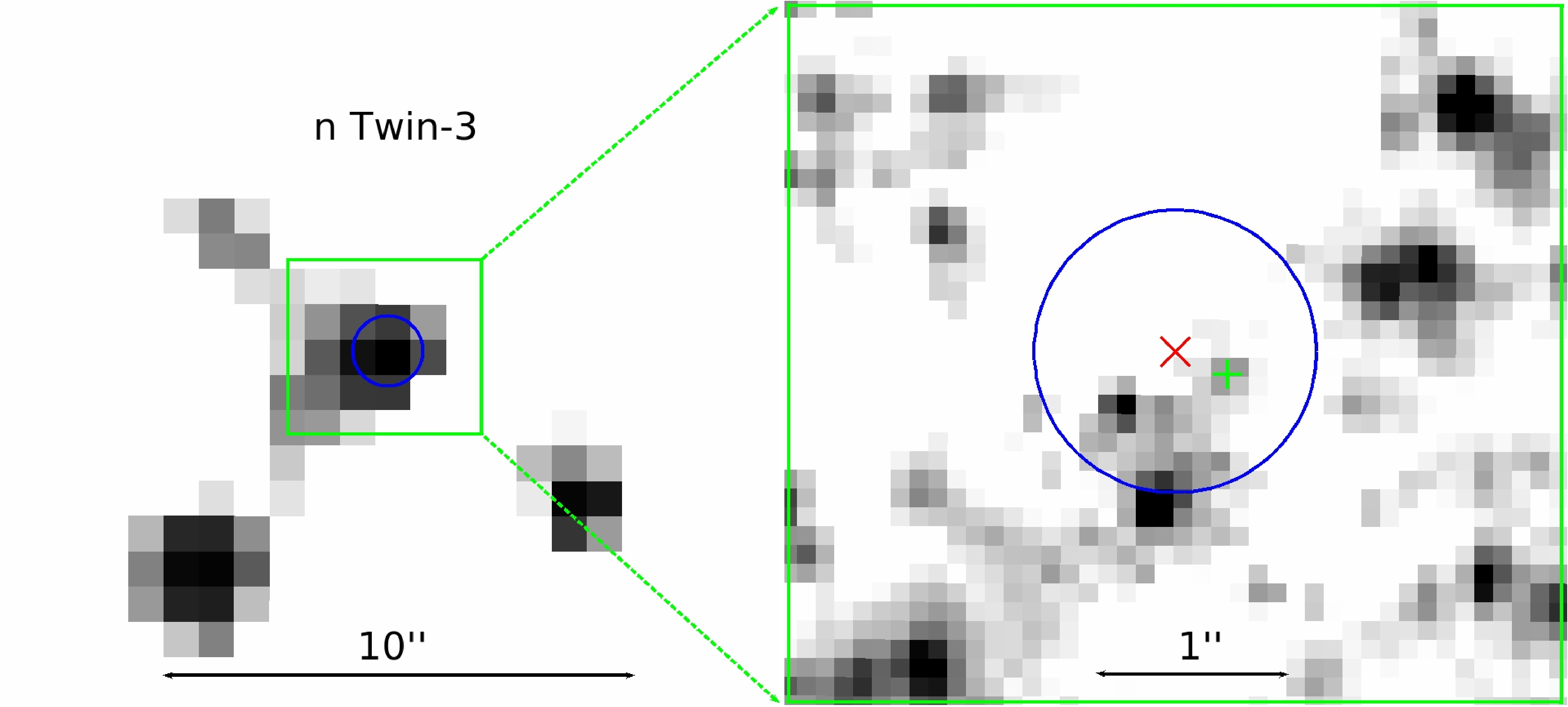} \\
\includegraphics[width=80mm]{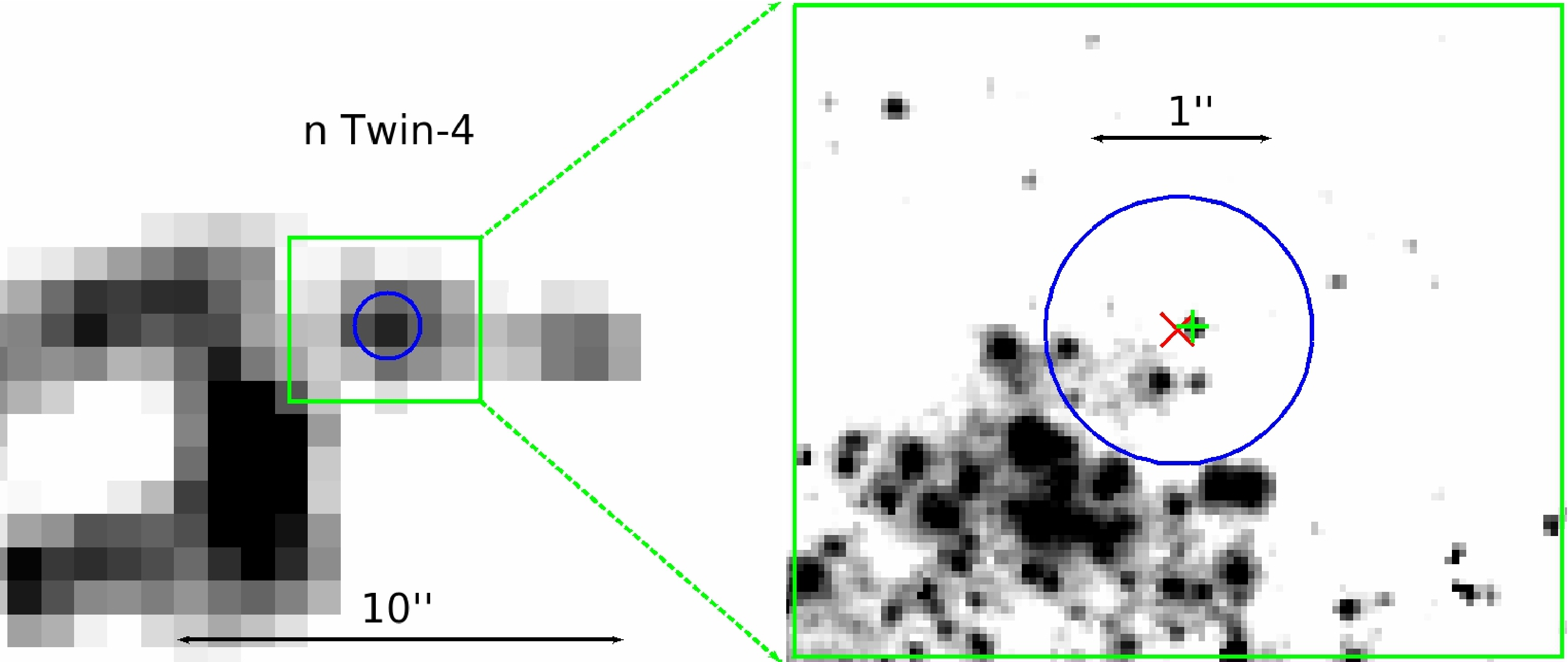} \\
\includegraphics[width=80mm]{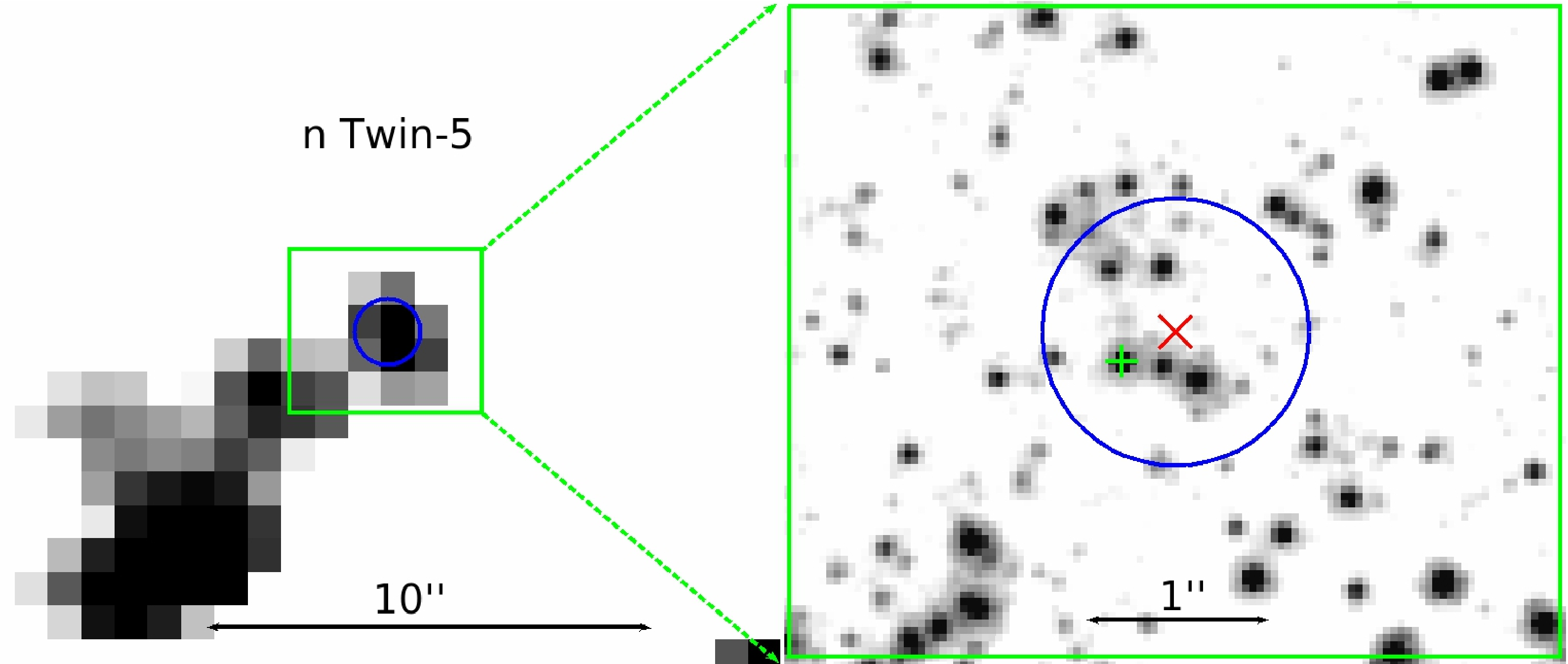}
\end{tabular}
\end{center}
\caption{\textit{Spitzer} IRAC\,$3.6\,\micron$ (left column) and HST $I$-band (F814W, right column) images of the candidate $\eta$\,Car analogs. 
The HST images were taken with ACS ($\eta$\,Twin-1), WFPC2 (2, 3) and WFC3 (4, 5). The circles 
are $1\farcs5$ in diameter, which is roughly equal to the IRAC\,$3.6\,\micron$ PSF FWHM. The rectangles on the left 
column enclose the regions shown on the right column. On the HST images, ``$\times$'' marks the location of the IRAC source (center of the circle) 
determined through pixel-to-pixel mapping and ``$+$'' marks the location of the likely optical counterpart as discussed in Section\,\ref{sec:data}.}
\label{fig:1}
\end{figure}

\begin{figure}[p]
\begin{center}
\includegraphics[width=160mm]{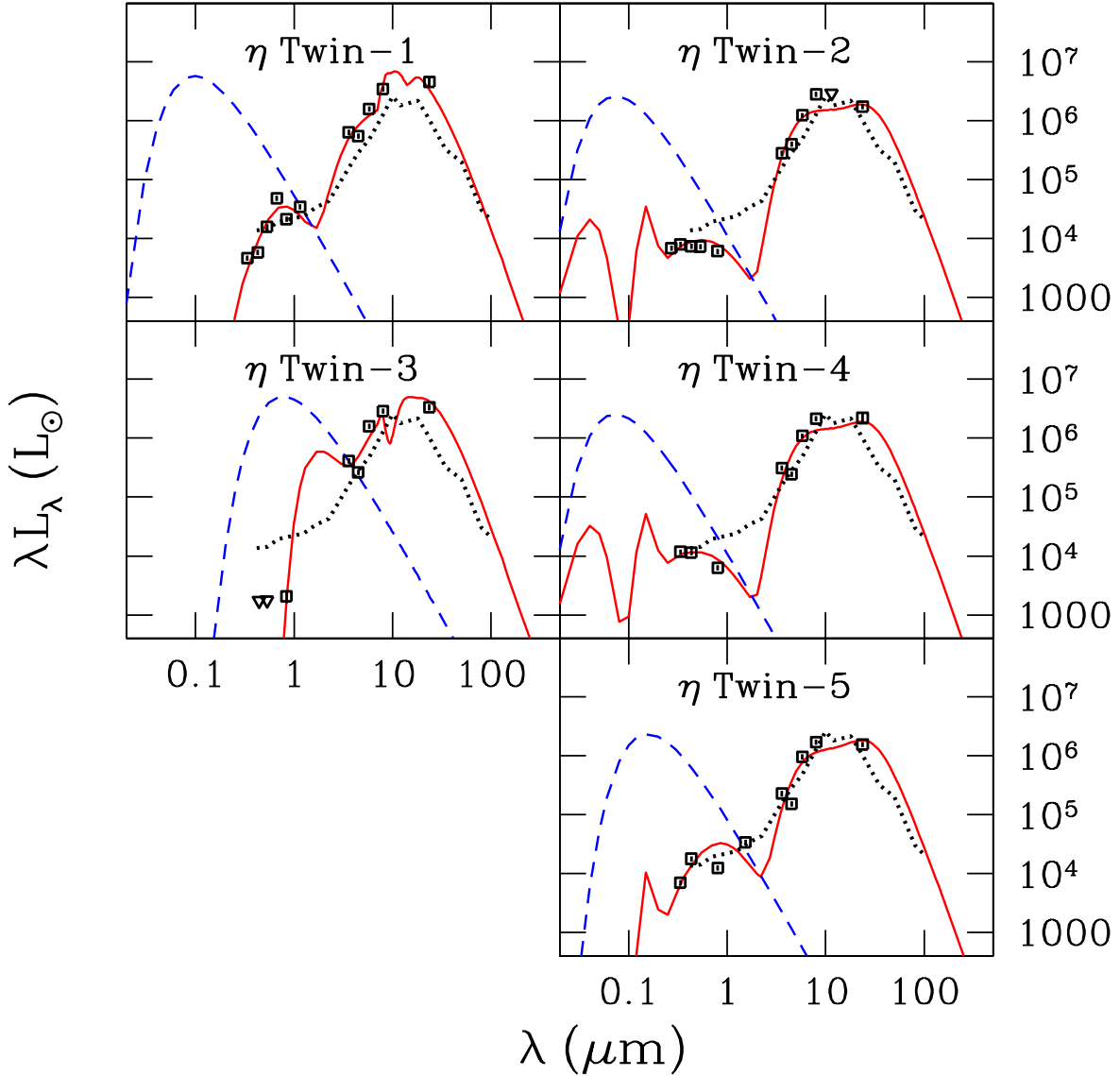}
\end{center}
\caption{The best fit models (solid line) for the observed SEDs (squares and triangles with the latter for 
upper limits) of the candidate $\eta$\,Car analogs and the SEDs of the underlying, unobscured blackbody sources (dashed line),
as compared to the SED of $\eta$\,Car (dotted line, from \citealp{ref:Robinson_1973}). The vertical bars show the larger of 
the flux uncertainties reported in Table\,\ref{tab:photo} and the $\sim10\%$ minimum flux uncertainty (0.1\,magnitude) used 
for SED modeling (not accounting for distance uncertainties). Here we show the best fit silicate models for $\eta$\,Twins-1 and 3, and
the best fit graphitic models for $\eta$\,Twins-2, 4 and 5.}
\label{fig:2}
\end{figure}

We identified five sources with mid-IR SEDs closely resembling that of $\eta$\,Car and optical fluxes or 
flux limits $\sim1.5-2$\,dex fainter than their mid-IR peaks. We will refer to these 
sources as $\eta$\,Twins-1, 2, 3, 4 and 5. We find one source each in M\,51 ($\eta$\,Twin-1), 
M\,101 ($\eta$\,Twin-2) and NGC\,6946 ($\eta$\,Twin-3), and two sources in M\,83 ($\eta$\,Twins-4, 5).
We identified HST counterparts of $\eta$\,Twins-1, 2 and 4 within the $0\farcs3$ matching radius.
For $\eta$\,Twin-3, no HST source is cataloged within the matching radius, so
we visually identified the closest location of flux excess at $\sim0\farcs3$, and used 
simple aperture photometry techniques to measure the $I$-band flux and the 
$B$ and $V$ band flux upper limits. For $\eta$\,Twin-5, although a cataloged 
HST source exists within the $0\farcs3$ matching radius, we selected a different 
source at $0\farcs35$ as the more likely photometric match because it is also a
bright HST $J$-band source. Table\,\ref{tab:photo} lists the photometry
of these sources, Figure\,\ref{fig:1} shows their IRAC\,$3.6\,\micron$ 
and HST $I$-band images, and Figure\,\ref{fig:2} shows their SEDs.
$\eta$\,Twins-1, 4 and 5 are H$\alpha$ emitters and $\eta$\,Twin-2 is a UV source (see Table\,\ref{tab:photo}).

We have  $UBVR$ variability data for M\,51, M\,101 and NGC\,6946
from the LBC survey for failed supernovae 
\citep{ref:Gerke_2015}. 
We analyzed 21/26/37~epochs of data 
spanning a 7.1/7.2/8~year period
for M\,51/M\,101/NGC\,6946 
with the ISIS image subtraction 
package~\citep{ref:Alard_1998}. We did not detect
any significant optical variability at the locations of $\eta$\,Twins-1, 2 or 3.

\citet{ref:Cutri_2012} identified $\eta$\,Twin-2
as a WISE point source and we use their $12\,\micron$ flux measurement 
as an upper limit for SED models (Section\,\ref{sec:analysis}). 
\citet{ref:Johnson_2001} reports an optically thick free-free radio source located 
$0\farcs49$ from $\eta$\,Twin-3 
and~\citet{ref:Hadfield_2005} identified a 
source with Wolf-Rayet spectroscopic signature $1\farcs54$ from $\eta$\,Twin-4.
We could not confirm if these sources are reasonable
astrometric matches to the IRAC locations.
$\eta$\,Twins-4 and 5 were cataloged by~\citet{ref:Williams_2015}
but not flagged as massive stars. 

\section{SED Modeling}
\label{sec:analysis}

We fit the SEDs of these five sources
using DUSTY~\citep{ref:Ivezic_1997} to model radiation 
transfer through a spherical medium surrounding a blackbody source, which is also a good 
approximation for a combination of unresolved non-spherical/patchy/multiple circumstellar shells.
We considered models with either 
graphitic or silicate dust~\citep{ref:Draine_1984}.
The models are defined by the stellar luminosity ($L_*$), 
stellar temperature ($T_*$),
$V$-band optical depth ($\tau_V$), dust 
temperature at the inner-edge of the shell ($T_d$) and shell 
thickness $\zeta=R_{out}/R_{in}$. 
We embedded DUSTY inside a Markov 
Chain Monte Carlo (MCMC) driver to fit each SED by varying 
$T_*$, $\tau_V$, and $T_d$ with $L_*$ determined by a $\chi^2$ fit 
for each model. We fix $\zeta=4$ since its exact value has little 
effect on the results \citep{ref:Khan_2015}, limit $T_*$ to a maximum value of 
$\sim50,000\,$K, set the minimum flux uncertainty to $\sim10\%$ (0.1\,magnitude) 
and do not account for distance uncertainties.

The best fit model parameters 
determine the radius of the inner edge of
the stellar-ejecta distribution ($R_{in}$).
The mass of the shell is $M_e = {4 \pi R_{in}^2 \tau_V}/{\kappa_V}$ 
(scaled to a visual opacity of $\kappa_V=100\,\kappa_{v100}$\,cm$^{2}/$g) and 
the age estimate for the shell is $t_e = {R_{in}}/{v_e}$ 
(scaled as $v_e=100\,v_{e100}$\,km\,s$^{-1}$) 
where we can ignore $R_{out}$ to zeroth order.
Table\,\ref{tab:mcmc} reports the parameters of the best fit models
and Figure\,\ref{fig:2} shows these models.
The integrated luminosity estimates depend little on 
the choice of dust type, and are in the range of $L_*\simeq10^{6.5-6.9}L_\odot$.
We also fit the SEDs using \citet{ref:Kurucz_2004} stellar atmosphere models instead
of blackbodies. Since these resulted in similar parameter estimates, we only report the 
blackbody results. 

Generally, the best fits derived for graphitic dust require lower optical 
depths, lower dust temperatures and larger shell radii, leading to higher 
ejecta masses and age estimates.
For $\eta$\,Twins-2 and 4, the stellar temperature estimates reach the 
allowed maximum of $\sim50,000\,$K. The best fit models of $\eta$\,Twin-1 and 5 
also require the presence of a hot star, but with temperatures lower than the allowed maximum
($\sim27,600/37,750$\,K and $\sim23,500/37,500$\,K for graphitic/silicate dust). 
Constrained by the low optical flux, the best fit models of $\eta$\,Twin-3 
require the presence of a cool star ($\sim5,000\,$K). 
For $\eta$\,Twins-2, 4 and 5,
the best fits derived for graphitic dust had lower $\chi^2$, and for 
for $\eta$\,Twins-1 and 3 the best fits derived for silicate dust have lower $\chi^2$.
Considering these models, the $\eta$\,Car analog candidates
appear to be embedded in $\sim5-10\,M_\odot$ of warm ($\sim400-600$\,K) obscuring material ejected a few centuries ago.

\begin{figure}[p]
\begin{center}
\includegraphics[width=100mm]{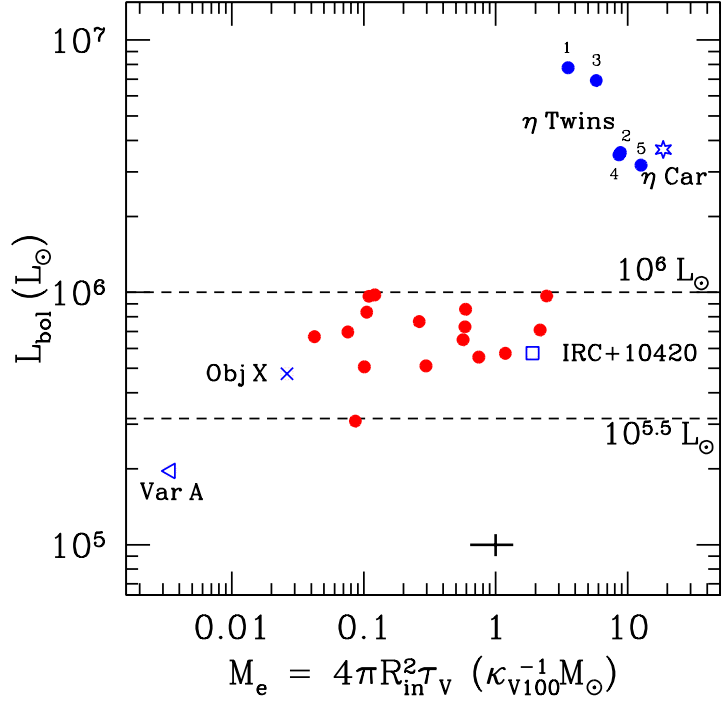}
\end{center}
\caption{The luminosities, $L_{bol}$, of the candidate $\eta$\,Car analogs (blue circles) 
as a function of the ejecta mass estimates, $M_e = {4 \pi R_{in}^2 \tau_V}/{\kappa_V}$, 
compared to the less luminous (red circles) 
self-obscured star candidates we identified in~\citet{ref:Khan_2015}.
The Galactic OH/IR star IRC+10420~\citep[e.g.,][square]{ref:Tiffany_2010}, M\,33's Variable\,A~\citep[e.g.,][triangle]{ref:Humphreys_1987},
Object\,X (``$\times$'',~\citealp{ref:Khan_2011}) and $\eta$\,Car (star symbol) 
are also shown for comparison. The error bar 
corresponds to the typical $1\sigma$ uncertainties on $L_{bol}$ ($\pm 10\%$) 
and $M_e$ ($\pm 35 \%$) of the best SED fit models.}
\label{fig:3}
\end{figure}

Figure\,\ref{fig:3} contrasts the bolometric luminosities and ejecta mass estimates of these 
five objects with the relatively less luminous sources we identified in~\citet{ref:Khan_2015}.
The five new sources form a distinct cluster 
close to $\eta$\,Car in the $L_{bol}-M_{ejecta}$ parameter space, whereas the previously identified dusty-star candidates
from~\citet{ref:Khan_2015} are more similar to the Galactic OH/IR star IRC+10420~\citep[e.g.,][]{ref:Tiffany_2010} 
or M\,33's Variable\,A~\citep[e.g.,][]{ref:Humphreys_1987}.

\section{Discussion}
\label{sec:discussion}

To an extragalactic observer located in one of the targeted galaxies surveying 
the Milky Way with telescopes similar to the HST and \textit{Spitzer}, 
$\eta$\,Car's present day SED would appear nearly identical to
the extragalactic $\eta$\,Car analog candidates we found. 
The Carina nebula is $\sim 2.5\arcdeg$ in extent \citep{ref:Smith_2007a} 
corresponding to $\sim 2\farcs5$ at our most distant galaxy (M\,51 at 8\,Mpc).
While this would not be resolved by \textit{Spitzer},
it would be easily resolved by HST.
Because more compact clusters are not uncommon,
in~\citet{ref:Khan_2013} we considered 
whether dusty clusters can hide $\eta$\,Car like stars and if 
we would confuse unresolved star-clusters with $\eta$\,Car analogs.
In general, a cluster sufficiently 
luminous to hide an evolved $\gtrsim30\,M_\odot$ star has hosted many luminous stars
with strong UV radiation fields and winds, which will generally clear the
cluster of the gas and dust needed to produce strong mid-IR emission
over the timescale that even the most massive star needs
to evolve away from the main sequence.
Moreover, emission from warm circumstellar ejecta 
peaks between the IRAC\,$8\,\micron$ and MIPS\,$24\,\micron$ bands 
and then turns over, as seen in all of our candidates, unlike
emission from colder intra-cluster dust that generally peaks
at longer wavelengths. 

A significant majority of massive stars are 
expected to be in multiple-star systems~\citep[e.g.,][]{ref:Sana_2011}, as is 
the case for $\eta$\,Car~\citep[e.g.,][]{ref:Damineli_1996,ref:Mehner_2010}. This is a 
minor complication, affecting luminosity estimates by at most a factor of $2$, and mass 
estimates even less. Assuming all the candidates we have identified are real analogs of $\eta$\,Car, then 
our galaxy sample (including the Milky Way) contains $N_c=6$ $\eta$\,Car-like stars.
Based on the ratio of star formation rates ($2$ vs. $10 M_\odot / $yr), our original 
sample of $7$ galaxies would be expected to have $\sim1$ $\eta$\,Car analog, which
is statistically consistent with not finding one in~\citet{ref:Khan_2015}.

If we expand our simple rate estimates from \citet{ref:Khan_2015},
our $N_c=6$ sources implies an eruption rate over the $12$ galaxies ($7$ previous, $4$ 
in this work, and the Milky Way)
of $F_e = 0.033 t_{d200}^{-1}$yr$^{-1}$ 
($0.016$yr$^{-1} < F_e t_{d200} < 0.059\,$yr$^{-1}$ at 90\% confidence)
where $t_d\simeq200$\,yrs is a rough estimate of the period over which our method 
would detect an $\eta$\,Car-like source. For comparison, the number of ccSN 
recorded in these galaxies over the past 30\,years is $10$ \citep[mainly based on][]{ref:Botticella_2012} 
for an SN rate of $F_{SN}=0.33/$yr.
This implies that the rate of $\eta$\,Car-like events is a fraction 
$f=0.094$ ($0.040 < f < 0.21$ at 90\% confidence) of the ccSNe rate.

If there is only one eruption mechanism and the SLSN-II are due to ccSN 
occurring inside these dense shells, then the ratio of the rate of $\eta$ Car-like 
events and SLSN-II, $r_{SLSN}/r_\eta = t_{SLSN}/t_\eta$ , is the ratio of the 
time period $t_{SLSN}$ during which the shell is close enough to the star to cause 
a SLSN to the time period $t_\eta$ over which shell ejections occur. With $r_{SLSN} \sim 10^{-3}$
of the core-collapse rate \citep{ref:Quimby_2013}, we must have that $t_{SLSN}/t_\eta \sim 10^{-2}$.
A typical estimate is that $t_{SLSN}\sim 10$ to $10^2$ years, which implies $t_\eta \sim 10^3$
to $10^4$~years, consistent with the properties of the massive shells around luminous stars
observed in our own Galaxy and suggesting that the instabilities driving the eruptions are linked to the onset
of carbon burning \citep{ref:Kochanek_2011}. This would also imply the existence of ``superluminous''
X-ray ccSN, where an older shell of material is too distant and low density to thermalize
the shock heated material but is still dense enough for the cooling time to be
faster than the expansion time. Such events should be $\sim 10$ times more common than
optical SLSN-II. If the eruptions driving SLSN-II are only associated
with later and shorter burning phases~\citep[e.g., as in][]{ref:Shiode_2014}
then there must be two eruption mechanisms and the vast majority of $\eta$\,Car analogs
will not be associated with the SLSN-II mechanism.

We identified the $5$ potential $\eta$\,Car analogs by
specifically focusing on finding sources that most closely resemble the SED of 
present day $\eta$\,Car. The reason that the SEDs of these five sources are so remarkably similar 
to each other is by design. We have not closely studied the less luminous mid-IR sources that may 
belong to the class of candidate self-obscured stars we identified in~\citet{ref:Khan_2015}, 
and some of the sources that we excluded because they have relatively bright optical 
counterparts may be evolved high mass stars with older, lower optical-depth shells.
It is readily apparent that a closer scrutiny 
of our mid-IR catalogs should reveal richer and more diverse 
populations of evolved massive stars. This in turn will let us better quantify the abundance of those stars, and 
constrain the rates of mass ejection episodes and mass loss from massive stars prior to their death by core-collapse.

The $\eta$\,Car analog candidates we identified can be studied at greater detail with the James Webb Space Telescope 
\citep[JWST, e.g.,][]{ref:Gardner_2006}, taking advantage of its order-of-magnitude-higher 
spatial resolution. 
These sources are luminous in the $3.6-24\,\micron$ wavelength range where the 
JWST will be most sensitive. They are rare laboratories for stellar astrophysics and will be very interesting 
extragalactic stellar targets for spectroscopic study with JWST's mid-IR instrument~\citep[MIRI,][]{ref:Rieke_2015}. This will 
give us an unprecedented view of these most-massive self-obscured stars, letting us study their evolutionary 
state and the composition of their circumstellar ejecta.

\acknowledgments
We thank the referee for helpful comments.
RK is supported by a JWST Fellowship awarded through the NASA 
Postdoctoral Program. SMA is supported by a Presidential Fellowship at The Ohio 
State University. KZS is supported in part by NSF Grant AST-151592.
CSK is supported by NSF grant AST-1515876.
This research has made use of
observations made with the \textit{Spitzer} Space Telescope, 
which is operated by the JPL and Caltech under a contract with NASA;
observations made with the NASA/ESA Hubble Space Telescope
and obtained from the Hubble Legacy Archive, which is a collaboration 
between the STScI/NASA, ST-ECF/ESA and the CADC/NRC/CSA; and the 
VizieR catalog access tool, CDS, Strasbourg, France.

\clearpage

\begin{table*}[p]   
\begin{center}     
\caption{Multi-Wavelength Photometry} 
\label{tab:photo}   
\begin{tabular}{rrrrrr}   
\\   
\hline  
\hline  
\\
\multicolumn{1}{c}{} &  
\multicolumn{1}{c}{$\eta$\,Twin-1} & 
\multicolumn{1}{c}{$\eta$\,Twin-2} & 
\multicolumn{1}{c}{$\eta$\,Twin-3} & 
\multicolumn{1}{c}{$\eta$\,Twin-4} & 
\multicolumn{1}{c}{$\eta$\,Twin-5} \\
\\
\hline  
\hline
\\
Host             &  M\,51          & M\,101          &   NGC\,6946      & M\,83          & M\,83          \\
RA (deg)         &  $202.46287$    & $210.80203$     &   $308.76548$    & $204.21782$    & $204.21523$    \\
Dec (deg)        &   $47.21126$    &  $54.31891$     &    $60.18314$    & $-29.88722$    & $-29.87481$    \\
$m_{UV}$         &  \dots          & $23.25\pm0.05$  &   \dots          & \dots          & \dots          \\
$m_U$            & $24.34$         & $23.16\pm0.03$  &   \dots          & $22.27\pm0.02$ & $22.84\pm0.04$ \\
$m_B$            & $25.14$         & $24.26\pm0.03$  & $>26.70$         & $23.31\pm0.02$ & $22.83\pm0.02$ \\
$m_V$            & $23.64\pm0.06$  & $23.92\pm0.03$  & $>26.15$         & \dots          & \dots          \\ 
$m_{H\alpha}$    & $18.26\pm0.01$  & \dots           &   \dots          & $22.00\pm0.06$ & $22.14\pm0.07$ \\ 
$m_R$            & $21.90\pm0.03$  & \dots           &   \dots          & \dots          & \dots          \\
$m_I$            & $22.31\pm0.08$  & $23.23\pm0.03$  & $24.56\pm0.26$   & $22.58\pm0.03$ & $21.84\pm0.02$ \\
$m_J$            & $21.12\pm0.07$  & \dots           &   \dots          & \dots          & \dots          \\
$m_H$            &  \dots          & \dots           &   \dots          & \dots          & $19.14\pm0.02$ \\
$m_{3.6}$        & $14.68\pm0.12$  & $15.08\pm0.11$  & $14.45\pm0.11$   & $14.26\pm0.10$ & $14.59\pm0.12$ \\
$m_{4.5}$        & $14.10\pm0.06$  & $13.97\pm0.04$  & $14.20\pm0.06$   & $13.79\pm0.07$ & $14.30\pm0.09$ \\
$m_{5.8}$        & $12.20\pm0.05$  & $11.99\pm0.06$  & $11.47\pm0.09$   & $11.40\pm0.08$ & $11.54\pm0.09$ \\
$m_{8.0}$        & $10.38\pm0.10$  & $10.12\pm0.01$  &  $9.84\pm0.06$   &  $9.70\pm0.11$ &  $9.95\pm0.07$ \\
$m_{12}$         & \dots           & $>8.81       $  &   \dots          & \dots          & \dots          \\
$m_{24}$         &  $6.50\pm0.20$  &  $7.07\pm0.02$  &  $6.09\pm0.13$   &  $6.09\pm0.20$ &  $6.47\pm0.05$ \\
\\
\hline  
\hline  
\end{tabular} 
\end{center}  
\tablenotetext{}{Vega\,magnitudes of the sources where the specific HST filters are
UV:\,F275W; $U$:\,F336W, $B$:\,F435W\,($\eta$\,Twin-2) / F438W\,(4,\,5) / F439W\,(1) / F450W\,(3); 
$V$:\,F555W, $H\alpha$:\,F656N\,($\eta$\,Twin-1) / F657N\,(4,\,5), $R$:\,F675W, $I$:\,F814W, 
$J$:\,F110W, $H$:\,F160W.
The HLA catalog reports no uncertainties for the U/B magnitudes of 
$\eta$\,Twin-1. The HST data sources are: $\eta$\,Twin-1: Prop.\,ID $7375$ (PI:\,Scoville) and Prop.\,ID $12490$ (PI:\,Koda);
$\eta$\,Twin-2: Prop.\,ID $9490$ (PI:\,Kuntz) and Prop.\,ID $13364$ (PI:\,Calzetti); 
$\eta$\,Twin-3: Prop.\,ID $9073$ (PI:\,Bregman); 
$\eta$\,Twins-4, 5: Prop.\,ID $12513$ (PI:\,Blair)}.
\end{table*}

\clearpage

\clearpage

\begin{table*}[p]   
\begin{center}   
\caption{Best Fit SED Models}  
\label{tab:mcmc}   
\begin{tabular}{rrrrrrrrrrrrrrrrrrrrr}   
\\   
\hline  
\hline    
\\   
\multicolumn{1}{c}{ID} &  
\multicolumn{1}{c}{} &
\multicolumn{1}{c}{} &
\multicolumn{1}{c}{$\chi^2/(m+n)$} & 
\multicolumn{1}{c}{$\tau_V$} & 
\multicolumn{1}{c}{$T_d$} & 
\multicolumn{1}{c}{$T_*$} & 
\multicolumn{1}{c}{$\log\left(R_{in}\right)$} & 
\multicolumn{1}{c}{$\log L_{bol}$} &
\multicolumn{1}{c}{$M_e$} &  
\multicolumn{1}{c}{$t_e$} 
\\   
\multicolumn{5}{c}{} &  
\multicolumn{1}{c}{(K)} & 
\multicolumn{1}{c}{(K)} & 
\multicolumn{1}{c}{(cm)} &
\multicolumn{1}{c}{($L_\odot$)} &
\multicolumn{1}{c}{($M_\odot$)}  & 
\multicolumn{1}{c}{(years)}
\\   
\hline  
\hline
\multicolumn{3}{c}{} &  
\multicolumn{8}{c}{\textit{Graphitic}} \\
\hline
\\
$\eta$\,Twin-1  & & & $ 151 / (11+0) $ & $  4.34  $ & $ 386 $ & $ 27,610 $ & $ 17.53 $ & $ 6.842 $ & $ 32.04 $ & $ 1087 $ \\
$\eta$\,Twin-2  & & & $  92 / (10+1) $ & $  2.51  $ & $ 404 $ & $ 50,120 $ & $ 17.37 $ & $ 6.554 $ & $  8.81 $ & $  750 $ \\
$\eta$\,Twin-3  & & & $  87 / (6+2)  $ & $ 16.67  $ & $ 337 $ & $  5,570 $ & $ 17.44 $ & $ 6.920 $ & $ 78.76 $ & $  870 $ \\
$\eta$\,Twin-4  & & & $ 113 / (8+0)  $ & $  2.24  $ & $ 394 $ & $ 50,110 $ & $ 17.39 $ & $ 6.544 $ & $  8.59 $ & $  783 $ \\
$\eta$\,Twin-5  & & & $ 269 / (9+0)  $ & $  3.61  $ & $ 375 $ & $ 23,530 $ & $ 17.37 $ & $ 6.504 $ & $ 12.62 $ & $  748 $ \\
\\
\hline
\multicolumn{3}{c}{} &
\multicolumn{8}{c}{\textit{Silicate}} \\
\hline
\\
$\eta$\,Twin-1  & & & $ 130 / (11+0) $ & $  6.34  $ & $ 603 $ & $ 37,740 $ & $ 16.97 $ & $ 6.890 $ & $  3.53 $ & $  299 $ \\
$\eta$\,Twin-2  & & & $ 205 / (10+1) $ & $  4.22  $ & $ 641 $ & $ 50,120 $ & $ 16.79 $ & $ 6.573 $ & $  0.99 $ & $  194 $ \\
$\eta$\,Twin-3  & & & $  69 / (6+2)  $ & $ 34.30  $ & $ 424 $ & $  4,730 $ & $ 16.71 $ & $ 6.839 $ & $  5.78 $ & $  164 $ \\
$\eta$\,Twin-4  & & & $ 136 / (8+0)  $ & $  3.94  $ & $ 622 $ & $ 50,120 $ & $ 16.80 $ & $ 6.555 $ & $  0.97 $ & $  199 $ \\
$\eta$\,Twin-5  & & & $ 307 / (9+0)  $ & $  4.57  $ & $ 760 $ & $ 37,520 $ & $ 16.59 $ & $ 6.449 $ & $  0.44 $ & $  124 $ \\
\\
\hline  
\hline  
\end{tabular} 
\end{center} 
\tablenotetext{}{The format $\chi^2/(m+n)$ indicates the goodness of fit $\chi^2$, 
the number of flux measurements $m$ used to determine the luminosity and the number of 
upper limits $n$ added to the estimate of $\chi^2$ once the luminosity is determined.}
\end{table*}

\clearpage

\clearpage

\end{document}